\documentclass{pasj00}
\draft

\newcommand{\actaa}{AcA, }

\begin{document}
\SetRunningHead{Author(s) in page-head}{Running Head}

\title{Multi-color photometric investigation of the totally eclipsing binary NO Cam}



%
 \author{Zhou \textsc{Xiao}\altaffilmark{1,2,3}
         Qian  \textsc{Shengbang}\altaffilmark{1,2,3,4}
         Zhang \textsc{Bin}\altaffilmark{1,2,3,4}}
\altaffiltext{1}{Yunnan Observatories, Chinese Academy of Sciences, 396 Yangfangwang, Guandu District, Kunming, 650216, P. R. China}
 \email{zhouxiaophy@ynao.ac.cn}
\altaffiltext{2}{Key Laboratory for the Structure and Evolution of Celestial Objects, Chinese Academy of Sciences, 396 Yangfangwang, Guandu District, Kunming, 650216, P. R. China}
\altaffiltext{3}{Center for Astronomical Mega-Science, Chinese Academy of Sciences, 20A Datun Road, Chaoyang District, Beijing, 100012, P. R. China}
\altaffiltext{4}{Graduate University of the Chinese Academy of Sciences, Yuquan Road 19, Sijingshang Block, 100049 Beijing, China}


\maketitle

\begin{abstract}
Multi-color photometric light curves of NO Cam in $V$ $R_C$ and $I_C$ bands are obtained and analyzed simultaneously using the Wilson-Devinney (W-D) program. The solutions suggest that NO Cam is an A-subtype overcontact binary with its mass ratio to be $q = 0.439$ and a contact degree of $f = 55.5\,\%$. The small temperature difference ($\Delta T = 44K$) between its two components indicates that the system is under thermal contact. The high orbital inclination ($i = 84.5^{\circ}$) strengthens our confidence in the parameters determined from the light curves. All available times of minimum light are collected and period variations are analyzed for the first time. The $O - C$ curve reveals that its period is increasing continuously at a rate of $dP/dt=+1.46\times{10^{-9}}$, which can be explained by mass transfer from the less massive component to the more massive one. After the upward parabolic variation is subtracted, the residuals suggest that there may be a cyclic variation with a period of 2.23 years and an amplitude of $A_3 = 0.00153$ days, which may due to the light-travel time effect (LTTE) arising from the gravitational influence of a close-in tertiary component. The close-in companion reveals that early dynamic interaction among triple system may have played very important role in the formation of the W UMa type binaries.
\end{abstract}

\section{Introduction}

The formation and evolutionary theory of W UMa type binaries is still an open issue. The most widely held view is that they are formed from initially detached binaries through nuclear evolution and angular momentum loss in the pre-contact phase. They will ultimately coalesce into single rapidly rotating stars, which may be progenitors of the poorly understood blue stragglers, FK Com-type stars, fast rotating A or F dwarf stars and so on \citep{1994ASPC...56..228B}. It has been demonstrated that the eruption of V1309 Sco was the result of a cool overcontact binary's merger \citep{2014ApJ...786...39N}. In 1970, \citet{1970VA.....12..217B} divided the W UMa type binaries into A-subtype and W-subtype. For A-subtype systems, the more massive component is the hotter one, while the less massive component is the hotter one in W-subtype systems.

Furthermore, more and more binaries are found in triple or multiple systems \citep{2010MNRAS.405.1930L} and several binaries with close-in companions have been reported recently \citep{2015ApJ...798L..42Q}, which makes the formation and evolutionary process of W UMa type binaries more complex. Therefore, detailed light curve and orbital period analysis on W UMa type binaries are still needed, which will provide invaluable information for the formation and evolutionary scenario of overcontact binaries.

NO Cam ( = NSV 1495, $V$ = $12^{m}.28$) is a newly discovered variable star observed by the ROTSE All-Sky Survey I \citep{2000AJ....119.1901A}. Then, it was listed as NSV 1495 in the NSV Catalogue \citep{2006IBVS.5699...26B} , pointing out that it was a W UMa type overcontact binary with spectral type to be F5. After that, several times of minimum light have been published (e.g., \citet{2006IBVS.5699...26B,2009IBVS.5894....1D,2010IBVS.5929....1N}). However, there isn't any spectroscopic element, photometric solution or period research. In the present paper, $V$ $R_C$ and $I_C$ bands light curves are analyzed using the Wilson-Devinney (W-D) program and reliable photometric parameters are obtained. All times of minimum light are collected and the period variations are determined. The dynamic interaction and evolutionary scenario are discussed to understand the nature of W UMa type binaries.

\section{Multi-color CCD Photometric Observations}

Multi-color light curves' observations were carried out on January 13 and March 15, 2013 using the 85cm reflecting telescope at Xinglong Observation Base, National Astronomical Observatories, Chinese Academy of Sciences. The telescope locates about 960 m above the sea-level. An Andor DW436 1K CCD camera is attached to the telescope. The effective field of view is 16.5 arcmin by 16.5 arcmin corresponding to a plate scale of 0.97 arcsec pixel$^{-1}$ \citep{2009RAA.....9..349Z}. The operating temperature of the CCD camera was set to be -55$^\circ$C. The broadband Johnson-Cousins $V$ $R_C$ $I_C$ filters were used during the observations. UCAC4 827-007726 and UCAC4 827-007718 were chosen as the comparison star (C) and the check star (Ch). Their coordinates and $V$ band magnitudes are listed in Table \ref{Coordinates1}. The finding chart is given as Fig. 1. The same integration time for observations on January 13 and March 15 were used, which were 30s for $V$ band, 20s for $R_C$ band, and 15s for $I_C$ band, respectively. Since the comparison and check stars are very close to the target, the extinction differences were negligible. The PHOT package in IRAF \footnote {The Image Reduction and Analysis Facility is hosted by the National Optical Astronomy Observatories in Tucson, Arizona at URL iraf.noao.edu.} was used to process the observational images. The average observational errors were 0.003 mag ($V$), 0.002 mag ($R_C$) and 0.002 mag ($I_C$), respectively. Differential aperture photometry method is applied to determine the light variations. The light curves are displayed in Fig. 2 and a few lines of the observational data are shown in Table 2. The standard deviations of the C - Ch data are all 0.008 mag for $V$, $R_C$, $I_C$ bands. The observational HJD are converted to phase with the following linear ephemeris:
\begin{equation}
Min.I(HJD)=2456306.2287+0^{d}.430754\times{E}.\label{linear ephemeris}
\end{equation}

\begin{figure}[!h]
\begin{center}
\includegraphics[width=10cm]{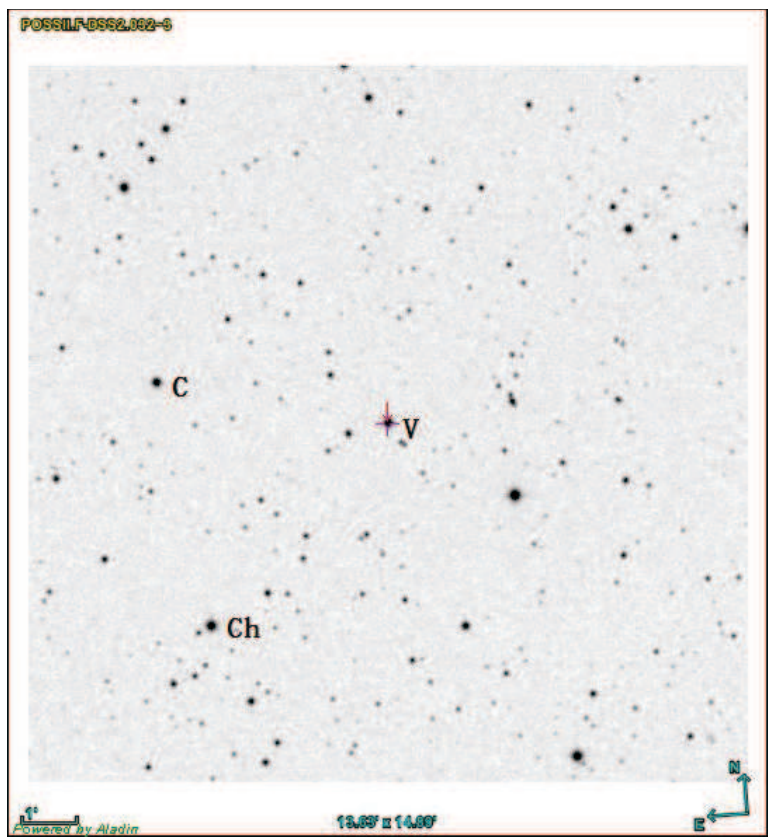}
\caption{The finding chart}
\end{center}
\end{figure}

\begin{table}[!h]
\begin{center}
\caption{Coordinates of NO Cam, UCAC4 827-007726, and UCAC4 827-007718.}\label{Coordinates1}
\begin{small}
\begin{tabular}{ccccc}\hline\hline
Targets          &   name               & $\alpha_{2000}$           &  $\delta_{2000}$         &  $V_{mag}$     \\ \hline
Variable         &   NO Cam             &$04^{h}14^{m}51^{s}.4$     & $+75^\circ20'40''.7$     &  $12.28$         \\
The comparison   &   UCAC4 827-007726   &$04^{h}15^{m}56^{s}.4$     & $+75^\circ21'40''.1$     &  $12.55$     \\
The check        &   UCAC4 827-007718	&$04^{h}15^{m}45^{s}.1$     & $+75^\circ17'14''.6$     &  $12.38$     \\
\hline\hline
\end{tabular}
\end{small}
\end{center}
\end{table}

\begin{figure}[!h]
\begin{center}
\includegraphics[width=13cm]{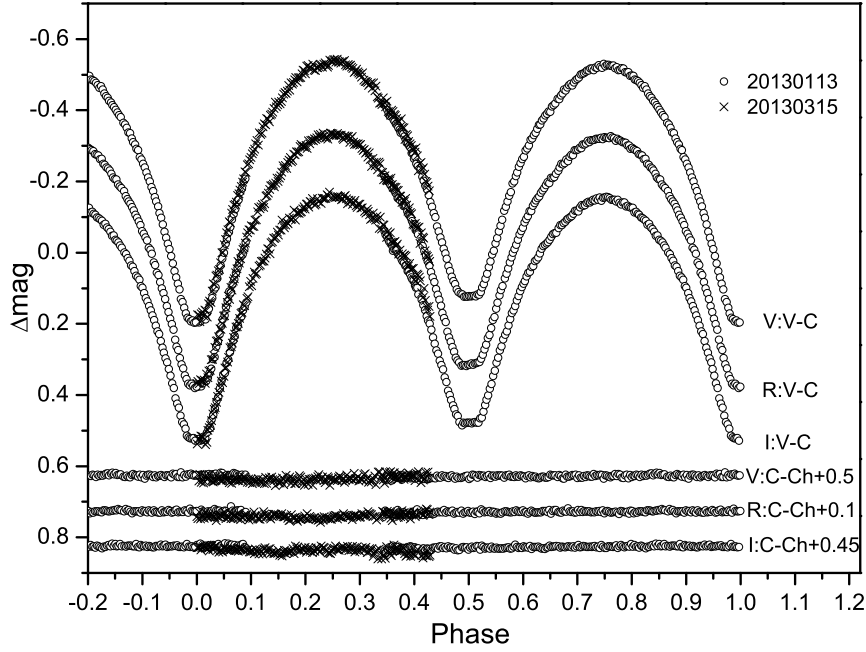}
\caption{Open circles and crosses refer to data observed on January 13 and March 15, respectively.}
\end{center}
\end{figure}

\begin{table}[!h]
\begin{center}
\caption{Observational data}\label{data}
\begin{small}
\begin{tabular}{ccccccccccccccc}\hline\hline
    HJD    &   Phase   &    $\Delta m$  &     HJD    &   Phase   &  $\Delta m$ &    HJD    &   Phase  &  $\Delta m$   \\
  2456300+ &           &                &  2456300+  &           &             & 2456300+  &          &            \\\hline
   5.9466  &  0.3451   &    -0.056      &    5.9563  &   0.3676  &  -0.009     &   5.9659  &  0.3899  &  0.050     \\
   5.9476  &  0.3474   &    -0.052      &    5.9573  &   0.3699  &   0.003     &   5.9668  &  0.3920  &  0.059     \\
   5.9485  &  0.3495   &    -0.049      &    5.9582  &   0.3720  &   0.006     &   5.9678  &  0.3943  &  0.065     \\
   5.9495  &  0.3518   &    -0.042      &    5.9592  &   0.3744  &   0.007     &   5.9688  &  0.3966  &  0.073     \\
   5.9505  &  0.3542   &    -0.037      &    5.9601  &   0.3764  &   0.016     &   5.9697  &  0.3987  &  0.073     \\
   5.9515  &  0.3565   &    -0.034      &    5.9611  &   0.3788  &   0.023     &   5.9707  &  0.4011  &  0.086     \\
   5.9525  &  0.3588   &    -0.019      &    5.9621  &   0.3811  &   0.026     &   5.9716  &  0.4031  &  0.097     \\
   5.9534  &  0.3609   &    -0.005      &    5.9630  &   0.3832  &   0.036     &   5.9726  &  0.4055  &  0.094     \\
   5.9544  &  0.3632   &    -0.009      &    5.9640  &   0.3855  &   0.035     &   5.9735  &  0.4076  &  0.110     \\
   5.9554  &  0.3655   &    -0.013      &    5.9649  &   0.3876  &   0.047     &   5.9760  &  0.4134  &  0.127     \\
\hline\hline
\end{tabular}
\end{small}
\end{center}
\textbf
{\footnotesize Notes.} \footnotesize These are only a few lines of the observational data; the whole data are available in the online version of this paper.
\end{table}

Meanwhile, six CCD times of minimum light were determined and the averaged values are listed in Table 3.

\begin{table}[!h]
\begin{center}
\caption{New CCD times of minimum light}\label{Newminimum}
\begin{tabular}{cccccc}\hline
    JD (Hel.)     &  Error (days)  & Min. &   Filter    &Telescopes\\\hline
  2456228.2601   & $\pm0.0008$   &   I  &   $BVR_CI_C$  &    60cm   \\
  2456301.2753   & $\pm0.0003$   &   II &   $VR_CI_C$   &    85cm   \\
  2456306.0135   & $\pm0.0001$   &   II &   $VR_CI_C$   &    85cm   \\
  2456306.2287   & $\pm0.0001$   &   I  &   $VR_CI_C$   &    85cm   \\
  2456351.0276   & $\pm0.0001$   &   I  &   $VR_CI_C$   &    85cm   \\
  2456630.1596   & $\pm0.0001$   &   I  &   $I_CN$      &    1m   \\

\hline
\end{tabular}
\end{center}
\textbf
{\footnotesize Notes.} \footnotesize 60cm and 1m denote to the 60cm and 1m reflecting telescope in Yunnan Observatories, and 85cm denotes to the 85cm reflecting telescope in Xinglong Observation base.
\end{table}

\section{Orbital Period Investigation}

O-C method is the traditional way to reveal variations on orbital period. In the present work, a total of 16 CCD times of minimum light are collected, as listed in Table 4. The lowercase p refers to the primary minimum and s refers to the secondary minimum. The initial residuals (Residuals$_1$) subtract from Equation (2) are listed in the fourth column of Table \ref{Minimum}.
\begin{equation}
Min.I(HJD)=2455480.8973+0^{d}.430754\times{E},\label{linear ephemeris}
\end{equation}
While a quadratic curve is applied to the initial residuals, Residuals$_2$ are obtained and a periodic variation appears, as displayed in the middle of Fig. 3. Therefore, a sinusoidal term is superimposed to the ephemeris. Based on the least-square method with the data weighted equally, the revised ephemeris is
\begin{equation}
\begin{array}{lll}
Min. I=2455480.8975(\pm0.0001)+0.43075666(\pm0.00000006)\times{E}
        \\+3.15(\pm0.08)\times{10^{-10}}\times{E^{2}}\\+0.00153(\pm0.00015)\sin[0.191(\pm0.001)\times{E}-20.9(\pm4.2)]
\end{array}
\end{equation}

It is conclude that the period is increasing continuously, at a rate of
$dP/dt=+1.46\times{10^{-9}}$, and the periodic change span is $P_3=2.22$ years with its amplitude to be $A_3 = 0.00153$ days. The results from equation (3) are displayed in Fig. 3.

\begin{table}[!h]
\caption{Times of minimum light}\label{Minimum}
\begin{center}
\small
\begin{tabular}{cclrrrcc}\hline\hline
  JD(Hel.)      &  p/s         &        Epoch        &     Residuals      &   Error   & Ref.  \\
 (2400000+)     &              &                     &                    &           &        \\\hline
51497.7180	    &  p	       &        -9247        &     0.00294        &           & 1      \\
54833.6883	    &  s	       &        -1502.5      &    -0.00111        &   0.0004  & 2      \\
55108.9387	    &  s	       &        -863.5	     &    -0.00252        &   0.0001  & 3      \\
55127.8930	    &  s	       &        -819.5	     &    -0.0014         &   0.0002  & 4      \\
55480.8973	    &  p	       &        0	         &     0              &   0.0002  & 5      \\
55590.3107	    &  p	       &        254	         &     0.00188        &   0.0020  & 6      \\
55590.5256	    &  s	       &        254.5	     &     0.00141        &   0.0016  & 6      \\
55592.2489	    &  s	       &        258.5	     &     0.00169        &   0.0010  & 6      \\
55599.3562	    &  p	       &        275	         &     0.00155        &   0.0008  & 6      \\
55599.5705	    &  s	       &        275.5	     &     0.00047        &   0.0014  & 6     \\
55627.3569	    &  p	       &        340	         &     0.00324        &   0.0047  & 7     \\
55627.5705	    &  s	       &        340.5	     &     0.00146        &   0.0042  & 7     \\
55670.4308	    &  p	       &        440	         &     0.00174        &   0.0037  & 7     \\
55863.8420	    &  p	       &        889	         &     0.00439        &   0.0004  & 7     \\
56228.2601	    &  p	       &        1735	     &     0.00461        &   0.0008  & 8     \\
56230.8428	    &  p	       &        1741	     &     0.00279        &   0.0019  & 9     \\
56301.2753	    &  s	       &        1904.5	     &     0.00701        &   0.0003  & 8     \\
56306.0135	    &  s	       &        1915.5	     &     0.00691        &   0.0001  & 8     \\
56306.2287	    &  p	       &        1916	     &     0.00674        &   0.0001  & 8     \\
56342.4110	    &  p	       &        2000	     &     0.0057         &   0.0008  & 10     \\
56351.0276	    &  p	       &        2020	     &     0.00722        &   0.0001  & 8     \\
56630.1596	    &  p	       &        2668	     &     0.01063        &   0.0001  & 8    \\\hline

\end{tabular}
\end{center}
\textbf
{\footnotesize Reference:} \footnotesize (1) \citet{2006IBVS.5699...26B}; (2) \citet{2009IBVS.5894....1D}; (3) \citet{2010IBVS.5929....1N}; (4) \citet{2010IBVS.5920....1D}; (5) \citet{2011IBVS.5960....1D}; (6) \citet{2011IBVS.5984....1H}; (7) \citet{2012IBVS.6010....1H}; (8) The present work; (9) \citet{2013IBVS.6042....1D}; (10) \citet{2013OEJV..160....1H}
\end{table}

\begin{figure}[!h]
\begin{center}
\includegraphics[width=13cm]{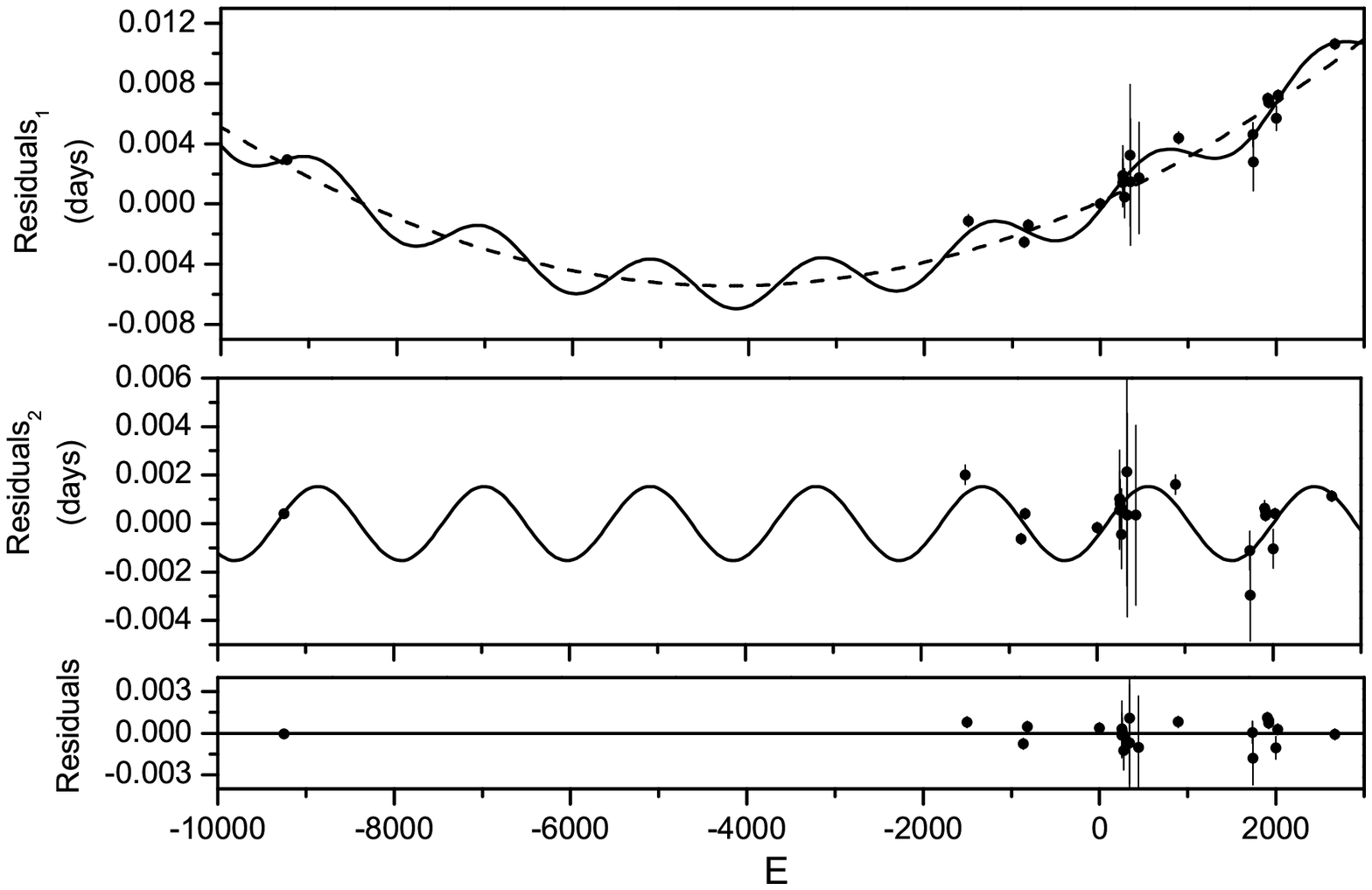}
\caption{In the upper panel, the dots represent the initial residuals calculated from Equation (2). The solid line is the theoretical curve corresponding to Equation (3), which contains a upward parabolic variation and a cyclic change, and the dash line refers to the quadratic term in the equation. In the middle panel, the quadratic part from Equation(3) is subtracted to make the periodic change more clear. The residuals are plotted in the bottom panel while both the upward parabolic change and the cyclic variation are removed.}
\end{center}
\end{figure}

Since the first data ( E = -9247 ) listed in Table 4 is quite far away from the others, we remove it and do another trial. The most probable ephemeris is:

\begin{equation}
\begin{array}{lll}
Min. I=2455480.8974(\pm0.0001)+0.4307565(\pm0.0000001)\times{E}
        \\+4.33(\pm0.58)\times{10^{-10}}\times{E^{2}}\\+0.00150(\pm0.00014)\sin[0.189(\pm0.001)\times{E}-17.0(\pm4.5)]
\end{array}
\end{equation}

The results also determined a continuous period increase ($dP/dt=+2.01\times{10^{-9}}$) and a cyclic period change ($P_3=2.25$ years, $A_3 = 0.00150$ days), as shown in Fig. 4.

Comparing the results from Equation (3) with Equation (4), we can conclude that the quadratic term do have a little difference, but the periodic part are almost the same. To check the reliability of the results, more times of minimum light are still needed in the future. We will adopt the results from Equation (3) for further discussion in the paper hereafter.

\begin{figure}[!h]
\begin{center}
\includegraphics[width=13cm]{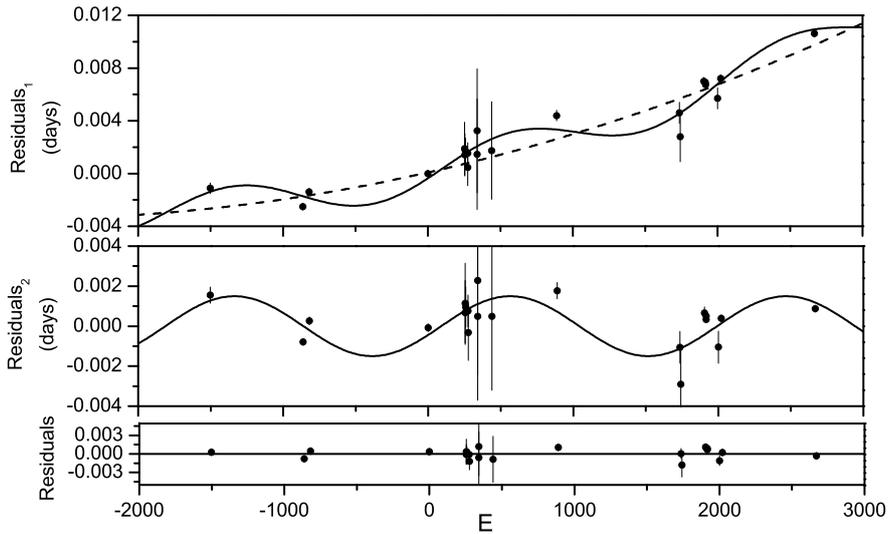}
\caption{The O - C curve from equation (4).}
\end{center}
\end{figure}

\section{Photometric Solutions}

As shown in Fig. 2, the light curves vary continuously, and the depth of the primary and secondary minima are almost equal, which indicate that NO Cam is a typical W UMa  type overcontact binary. To derive its physical parameters, the 2013 Version of Wilson-Devinney (W-D) program \citep{Wilson1971,Van2007,Wilson2010} is used. The number of observational data applied in the program are 444 in $V$ band, 445 in $R_C$ band, and 437 in $I_C$ band. The observational HJD were converted to phase while applying the program.

The spectral type given in the NSV Catalogue \citep{2006IBVS.5699...26B} is F5. The 2MASS All Sky Catalogue \citep{2003yCat.2246....0C} gives the color index of $J - H = 0.220$, which also corresponds to a spectral type of F5. Thus, the effective surface temperature of the primary star (the star eclipsed at primary minimum) is set to be $T_1 = 6530K$ \citep{Cox2000}. Convective atmospheres are assumed, and the corresponding gravity-darkening coefficients of $g_1=g_2=0.32$ \citep{1967ZA.....65...89L} and bolometric albedo of $A_1=A_2=0.5$ \citep{1969AcA....19..245R} are assigned. The limb darkening coefficients originate from \citet{1993AJ....106.2096V}'s table by considering logarithmic law accordingly.

Mode 3 for contact system is adopted. As it is a total eclipsing binary, the mass ratio will be well-determined from light curves since the amplitude of light variation depends mainly on the mass ratio for total-annular eclipsing binaries, whereas the amplitude also depends strongly on inclination\citep{2005Ap&SS.296..221T} for partial eclipses. The adjustable parameters are: the mass ratio $q$ ($M_2/M_1$); the orbital inclination $i$; the effective temperature of the secondary star ($T_{2}$); the monochromatic luminosity of the primary star ($L_{1V}$, $L_{1R}$ and $L_{1I}$); the dimensionless potential of the primary and secondary stars ($\Omega_{1}=\Omega_{2}$ in this case) and the third light ($l_3$). In our solutions, we find that the contribution of third light is negligible. The converged photometric solutions are listed in Table 5, Solutions A. The theoretical light curves are displayed in Fig. 5. The standard deviations of the fitting residuals is 0.009 mag ($V$), 0.007 mag ($R_C$) and 0.007 mag ($I_C$), respectively.

 Since the light curves show very good symmetry and Solution A for convective parameters imply that there may be no spot on the stars, radiative parameters of $g_1=g_2=1.00$ \citep{1967ZA.....65...89L} and $A_1=A_2=1.00$ \citep{1969AcA....19..245R} are tried, and Solutions B is obtained in Table 5. The theoretical light curves for Solutions B is almost the same with those in Fig. 5, so they are not displayed here. Comparing Solutions A with Solutions B, it is found that convective parameters determines a smaller temperature difference and residual, which may be much more acceptable. Solutions A will be referred for further discussion hereafter.

\begin{table}[!h]
\begin{center}
\caption{Photometric solutions}\label{phsolutions}
\small
\begin{tabular}{lllllllll}
\hline\hline
Parameters                            &  Solutions A                  &  Solutions B                   \\\hline
$T_{1}(K)   $                         &  6530(fixed)                  &  6530(fixed)                   \\
$g_{1}$                               &  0.32(fixed)                  &  1.00(fixed)                  \\
$g_{2}$                               &  0.32(fixed)                  &  1.00(fixed)                  \\
$A_{1}$                               &  0.50(fixed)                  &  1.00(fixed)                   \\
$A_{2}$                               &  0.50(fixed)                  &  1.00(fixed)                 \\
$a(R_\odot)$                          &  3.03(fixed)                  &  3.02(fixed)                 \\
q ($M_2/M_1$ )                        &  0.439($\pm0.001$)            &  0.416($\pm0.001$)             \\
$i(^{\circ})$                         &  84.5($\pm0.1$)               &  83.8($\pm0.1$)               \\
$\Omega_{1}=\Omega_{2}$               &  2.6099($\pm0.0036$)          &  2.6097($\pm0.0036$)         \\
$T_{2}(K)$                            &  6486($\pm3)$                 &  6304($\pm4$)                 \\
$\Delta T(K)$                         &  44                           &  226                            \\
$T_{2}/T_{1}$                         &  0.9933($\pm0.0005$)          &  0.9654($\pm0.0006$)          \\
$L_{1}/(L_{1}+L_{2}$) ($V$)           &  0.6731($\pm0.0002$)          &  0.7141($\pm0.0003$)           \\
$L_{1}/(L_{1}+L_{2}$) ($R_C$)         &  0.6721($\pm0.0003$)          &  0.7094($\pm0.0003$)           \\
$L_{1}/(L_{1}+L_{2}$) ($I_C$)         &  0.6713($\pm0.0003$)          &  0.7052($\pm0.0003$)          \\
$r_{1}(pole)$                         &  0.4527($\pm0.0005$)          &  0.4483($\pm0.0005$)          \\
$r_{1}(side)$                         &  0.4896($\pm0.0006$)          &  0.4830($\pm0.0006$)          \\
$r_{1}(back)$                         &  0.5298($\pm0.0008$)          &  0.5179($\pm0.0007$)          \\
$r_{2}(pole)$                         &  0.3189($\pm0.0015$)          &  0.3053($\pm0.0014$)           \\
$r_{2}(side)$                         &  0.3378($\pm0.0019$)          &  0.3216($\pm0.0018$)           \\
$r_{2}(back)$                         &  0.3986($\pm0.0044$)          &  0.3713($\pm0.0036$)           \\
$f$                                   &  $55.5\,\%$($\pm$1.4\,\%$$)   &  $39.6\,\%$($\pm$1.4\,\%$$)  \\
$\Sigma{\omega(O-C)^2}$               &  0.005229                     &  0.005288                     \\
\hline
\end{tabular}
\end{center}
\end{table}

\begin{figure}[!h]
\begin{center}
\includegraphics[width=13cm]{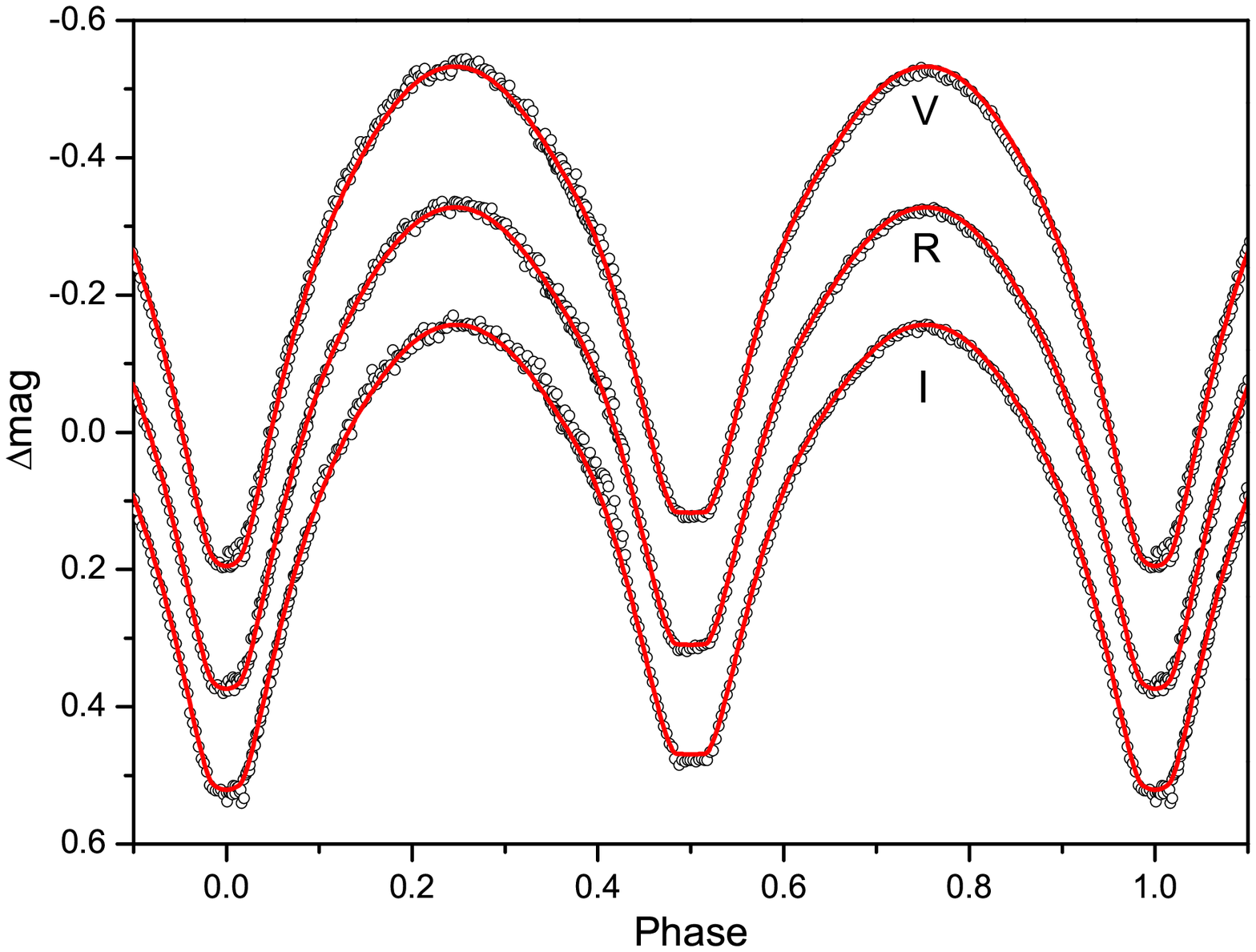}
\caption{Open circles refer to observed light curves and red lines correspond to theoretical light curves.}
\end{center}
\end{figure}

\section{Discussions and Conclusions}

The light curves' solutions suggest that NO Cam is an A-subtype overcontact binary with its mass ratio to be $q = 0.439$. The contact degree of $f = 55.5\,\%$ reveals that it is a deep contact system \citep{2005AJ....130..224Q}. The two components have nearly the same surface temperature ($\Delta T = 44K$) in spite of their quite different mass and radii, which indicates that the system is under thermal contact. The primary component contributes about $67\,\%$ of the total luminosity and the existence of the third light ($l_3$) are not detected. The orbital inclination is $i = 84.5^{\circ}$, which confirms our assumption that NO Cam is a totally eclipsing binary. It indicates that those photometric parameters are derived reliably. As there is no radial velocity curves, we cannot give a direct determination of absolute parameters. Assuming that the primary component is a normal main-sequence star, its mass is estimated to be $M_1 = 1.4M_\odot$ \citep{Cox2000}. Considering the mass ratio obtained, the mass of the secondary star is calculated to be $M_2 = 0.61M_\odot$. The semi-major axis listed in Table 5 are calculated according to the assumed masses. According to the statistical analysis conducted by
\citet{2015AJ....150...69Y}, F-type contact binaries have a really high ratio in extreme mass ratio, overcontact binary systems. It is supposed that NO Cam may evolve into a extreme mass ratio, overcontact binary system and merge into a single rapidly rotating star as in the case of V1309 Sco \citep{2014ApJ...786...39N,2016RAA....16d..16Z}.

The period analysis shows that its orbital period is increasing at a rate of $dP/dt=+1.46\times{10^{-9}}$. To account for the period increase, a conservative mass transfer mechanism is supposed. Applying the estimated masses to the well-known equation
\begin{equation}
\begin{array}{lll}
 \frac{dM_{2}}{dt}= -\frac{M_1M_2}{3P(M_1-M_2)}\times{dP/dt},
 \end{array}
\end{equation}
we conclude that the period increase is due to the mass transfer from the less massive component to the more massive one. The mass transfer rate is $\frac{dM_{2}}{dt}=8.37\times{10^{-8}}M_\odot/year$. However, the real mass transfer rate may be quite different from this value while the contribution of angular momentum loss (AML) is considered.

As shown in the middle panel of Fig. 3, there may be a cyclic variation superimposed in the period variations. Cyclic oscillations are usually encountered for W UMa-type overcontact binary stars \citep{2003A&A...400..649Q} which could be plausibly explained as the light-travel time effect arising from the gravitational influence of a third object \citep{2010MNRAS.405.1930L}. By assuming a circular orbit, the projected orbital radius rotating around the barycenter of the triple system is calculated with the following equation,
\begin{equation}
\begin{array}{lll}
a'_{12}\sin i'=A_3 \times c,
 \end{array}
\end{equation}
where $A_3$ is the amplitude of periodic variation and $c$ is the speed of light. Therefore, the projected orbital radius is calculated to be $a'_{12}\sin i'=0.26(\pm0.02)AU$. Considering the mass function of the triple system and the masses of the primary and secondary stars, the mass and orbital radius of the tertiary companion are computed with the following equation,
 \begin{equation}
\begin{array}{lll}
f(m)=\frac{4\pi^2}{GP^2_3}\times(a'_{12}\sin i')^3=\frac{(M_3\sin i')^3}{(M_1+M_2+M_3)^2},
 \end{array}
\end{equation}
where $G$ is the gravitational constant and $P_3$ is the period that the tertiary component orbits around NO Cam. The mass function are calculated to be $f(m) = 0.00373M_\odot$. The $i'$-$M_3$ and $i'$-$a_3$ diagrams are shown in Fig. 6. The red stars in Fig. 6 are the mass and orbital radius of the tertiary component when the three components are coplanar.

\begin{figure}[!h]
\begin{center}
\includegraphics[width=14cm]{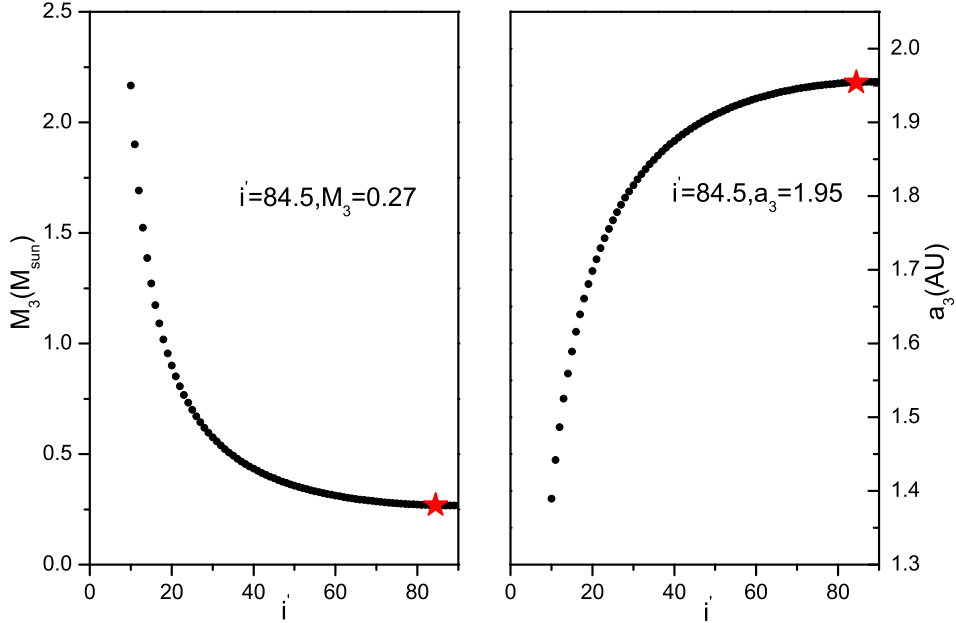}
\caption{Left panel: relations between the mass $M_3$ ( in units of $M_\odot$) of the third component and its orbital inclination $i'(^{\circ})$. Right panel: relations between the orbital radius $a_3$ ( in units of AU) of the third component and its orbital inclination $i'(^{\circ})$. The red star is the position of the tertiary component when it is coplanar with NO Cam.}
\end{center}
\end{figure}

Although tertiary components are quite common in W UMa systems, their effects to the hosting binaries are not clear. It is supposed to have accelerated the orbital evolution of the hosting binary by removing angular momentum from it during the early dynamical interaction \citep{2013ApJS..209...13Q,2014AJ....148...79Q}. According to our calculation, NO Cam is a triple system with a cool stellar companion at an orbital separation about 1.95 AU. Recently, some binaries with close-in companions have been reported as listed in Table 6. It has to be mentioned that the masses are minimum values since we do not know the orbital inclination ($i'$) of the tertiary components. The most special one is CSTAR 038663 in which the tertiary component is only about 1AU away from the central binary system. Dynamic interaction among the triple system must have significant influence on the formation of the binary system. Contact binaries with close-in companions are important targets for testing theories of star formation and interaction, and should be long term studied.

\begin{table}[!h]
\begin{center}
\caption{Parameters of the close-in companions to contact binaries ($i^{'}=90^{\circ}$)}
\begin{tabular}{lccccc}\hline
    Target              &  $M_3(M_\odot)$    &   $a_3(AU)$   &     Ref.\\\hline
  PY Vir                & $0.79$             &   $2.8$       &      1   \\
  V401 Cyg              & $0.70$             &   $2.4$       &      2   \\
  CSTAR 038663          & $0.63$             &   $0.93$      &      3   \\
  SDSS J001641-000925   & $0.14$             &   $2.8$       &      4   \\
  V384 Ser              & $0.37$             &   $2.0$       &      5   \\
  NO Cam                & $0.27$             &   $1.96$      &      6   \\
\hline
\end{tabular}
\end{center}
\textbf
{\footnotesize Reference:} \footnotesize (1) \citet{2013AJ....145...39Z};  (2) \citet{2013AJ....146...28Z}; (3) \citet{2014ApJS..212....4Q}; (4) \citet{2015ApJ...798L..42Q}; (5) \citet{2015PKAS...30..215L}; (6) The present work.
\end{table}

\bigskip

\vskip 0.3in \noindent
This work is supported by the Chinese Natural Science Foundation (Grant No. 11133007, 11325315, 11203066 and 11403095), the Strategic Priority Research Program ``The Emergence of Cosmological Structure'' of the Chinese Academy of Sciences (Grant No. XDB09010202) and the Science Foundation of Yunnan Province (Grant No. 2012HC011 and 2014FB187). New CCD photometric observations of NO Cam were obtained with the 60cm and 1.0m telescopes at Yunnan Observatories, and the 85cm telescope in Xinglong Observation base in China. This research has made use of the SIMBAD database, operated at CDS, Strasbourg, France and the fourth US Naval Observatory CCD Astrograph Catalog (UCAC4).


\end{document}